\documentclass[letterpaper]{article} 
\usepackage{aaai24}  
\usepackage{times}  
\usepackage{helvet}  
\usepackage{courier}  
\usepackage[hyphens]{url}  
\usepackage{graphicx} 
\usepackage{natbib}  
\usepackage{caption} 
\usepackage{multirow}

\usepackage{tabularx}
\usepackage{booktabs}
\usepackage{algorithm}
\usepackage{algorithmic}
\usepackage{xcolor}
\usepackage{amsmath}
\usepackage{comment}
\usepackage{newfloat}
\usepackage{listings}
\DeclareCaptionStyle{ruled}{labelfont=normalfont,labelsep=colon,strut=off}
\floatstyle{ruled}
\newfloat{listing}{tb}{lst}{}
\floatname{listing}{Listing}
\newcommand{\primarykeywords}{Urban delivery networks, minimum
  spanning trees, graph redundancy, resilient routing, order bundling,
multi-agent coordination}

\title{Multi-Agent Training-free Urban Food Delivery System using
Resilient UMST Network}
\author{
  Md Nahid Hasan\textsuperscript{\rm 1},
  Vishwam Tiwari\textsuperscript{\rm 2},
  Aditya Challa\textsuperscript{\rm 2},
  Vaskar Raychoudhury\textsuperscript{\rm 1},
  Snehanshu Saha\textsuperscript{\rm 2}
}
\affiliations{
  \textsuperscript{\rm 1}Department of Computer Science and Software
  Engineering, Miami University, Oxford, Ohio, USA\\
  \textsuperscript{\rm 2}Department of Computer Science \&
  Information Systems, BITS Pilani K. K. Birla Goa Campus, and
  APPCAIR, BITS Goa, India\\
  hasanm13@miamioh.edu,
  f20240591@goa.bits-pilani.ac.in,
  adityac@goa.bits-pilani.ac.in,
  raychov@miamioh.edu,
  snehanshus@goa.bits-pilani.ac.in
}

\usepackage{bibentry}

\begin{document}

\maketitle

\begin{abstract}

Delivery systems have become a core part of urban life, supporting the demand for food, medicine, and other goods. Yet traditional logistics networks remain fragile, often struggling to adapt to road closures, accidents, and shifting demand. Online Food Delivery (OFD) platforms now represent a cornerstone of urban logistics, with the global market projected to grow to over 500 billion USD
by 2030. Designing delivery networks that are efficient and resilient remains a major challenge: fully connected graphs provide flexibility but are computationally infeasible at scale, while single Minimum Spanning Trees (MSTs) are efficient but easily disrupted.

We propose the Union of Minimum Spanning Trees (UMST) approach to construct delivery networks that are sparse yet robust. UMST generates multiple MSTs through randomized edge perturbations and unites them, producing graphs with far fewer edges than fully connected networks while maintaining multiple alternative routes between delivery hotspots. 
Across multiple U.S. cities, UMST achieves 20--40$\times$ fewer edges than fully connected graphs while enabling substantial order bundling with 75--83\% participation rates. Compared to learning-based baselines including MADDPG and Graph Neural Networks, UMST delivers competitive performance (88-96\% success rates, 44-53\% distance savings) without requiring training, achieving 30$\times$ faster execution while maintaining interpretable routing structures. Its combination of structural efficiency and operational flexibility offers a scalable and resilient foundation for urban delivery networks.
\end{abstract}







\section{Introduction}
\label{sec:intro}

Online Food Delivery (OFD) has evolved from a convenience service into a core component of urban infrastructure. The global OFD market was valued at \$288.84 billion in 2024 and is projected to exceed \$505.50 billion by 2030 \cite{GV2025}, reflecting a sustained shift in consumer expectations around convenience, reliability, and on-demand service availability. As cities become more densely populated, delivery services increasingly operate as a parallel transportation ecosystem that must be optimized for efficiency, scalability, and resilience.

Despite rapid advances in OFD platforms, the underlying delivery networks remain very brittle. Many current systems rely on routing algorithms built on static, fully connected graphs or simplified network structures that fail to capture the spatial-temporal variability of real cities. In practice, delivery demand is highly non-uniform: commercial zones experience sharp midday surges, residential areas dominate during evening hours, and some regions exhibit consistently low demand. These variations place uneven load on the network and complicate efficient routing at scale.

Traditional graph-based algorithms such as Dijkstra's and Floyd-Warshall perform well for static or small-scale problems. However, urban OFD environments are highly dynamic: vehicle flows, road closures, and weather conditions can change suddenly, while demand surges vary across the city. Both fully connected and sparse networks limit the ability to bundle multiple deliveries efficiently, as single shortest paths in sparse graphs often do not provide the alternative routes needed to combine orders into shared trips.

Reinforcement Learning (RL) approaches have emerged as a promising method for urban delivery \cite{mehra2024deliverai,10705350}. However, RL-based methods still rely heavily on the underlying graph representation. Fully connected graphs provide maximal routing flexibility but require $O(|H|^2)$ edges, making them computationally intractable for city-scale deployments with hundreds of hotspots. Simplified structures, such as single MSTs, reduce computational load but sacrifice redundancy, leaving the network fragile to edge failures and limiting flexibility for multi-stop deliveries.

Recent research has incorporated urban structure explicitly through spatial decomposition and hierarchical clustering \cite{10705350}. While these approaches improve routing efficiency, most still rely on static graph designs, such as a single MST, which offer minimal alternative paths. Networks with limited redundancy are particularly vulnerable to road closures or temporary blockages; even small disruptions can disconnect key regions or force long detours.

\subsection{Motivation and Contribution}

To overcome these limitations, we introduce the Union of Minimum Spanning Trees (UMST), a principled method for constructing delivery networks that are both sparse and resilient. UMST builds multiple MSTs by applying randomized edge perturbations and then takes the union of the resulting trees. This process ensures that the network includes multiple independent paths between hotspots. By introducing redundancy during graph construction, UMST achieves a balance between computational efficiency and robustness without requiring adaptive routing at delivery time.

The resulting UMST network contains $O(k|H|)$ edges (for $k$ MSTs), dramatically fewer than the $O(|H|^2)$ edges of a fully connected graph, yet provides sufficient alternative routes to improve both shortest-path efficiency and multi-stop delivery flexibility. Across multiple U.S. cities, UMST consistently enables high order bundling participation (75 to 83\%), achieves strong success rates (88 to 96\%), and delivers substantial distance savings (44 to 53\%) compared to no-bundling baselines. Importantly, \textbf{UMST requires no training} and adapts immediately to network changes through simple graph reconstruction, offering 30 times faster execution than learning-based methods while maintaining competitive or superior performance.

Our comparison with learning-based approaches including MADDPG and Graph Neural Networks demonstrates that carefully designed graph structures can match or exceed the performance of complex learning systems while providing interpretable routing, zero training overhead, and immediate adaptability to disruptions. This makes UMST a robust foundation for multi-agent urban delivery systems, offering predictable performance even under structural disruptions.

Our main contributions can be summarized as follows:

\begin{itemize}

    \item \textbf{Building resilient delivery networks:} We introduce UMST, which combines multiple MSTs built on slightly perturbed versions of the network to create a sparse but robust delivery graph with $O(k|H|)$ edges.

    \item \textbf{Supporting efficient order bundling:} By providing multiple alternative paths between hotspots, UMST achieves 75 to 83\% bundling participation rates, consolidating deliveries into shared routes and reducing vehicle distance by 44 to 53\%.

    \item \textbf{Competitive performance without training:} UMST matches or outperforms learning-based baselines (MADDPG, GNN) in key metrics while requiring no training, executing 30 times faster, and providing interpretable routing structures suitable for real-time deployment.

    \item \textbf{Reliable performance under disruption:} UMST maintains high connectivity and route flexibility through its multi-path structure, ensuring the network continues to function when edges fail or demand patterns shift.

\end{itemize}


\section{Related Works}


In this section, we discuss some key techniques for online food delivery developed in recent times.

\noindent \textbf{Deterministic and Heuristic Approaches:} Prior heuristic solutions have leveraged existing transportation infrastructure to facilitate package deliveries. CrowdDeliver \cite{crowddeliver, liu2018foodnet, du2019crowdnet} and PPtaxi \cite{pptaxi, seng2023ridesharing} repurpose passenger-taxi fleets for parcel transport, utilizing idle vehicle capacity. Other heuristic models, such as MDMPMP \cite{mdmpmp}, apply time-expanded graph techniques to optimize multi-hop parcel synchronization. However, these systems often prioritize passenger transportation efficiency over dedicated logistics optimization, limiting their applicability to pure delivery contexts.

Recent work has explored a wide range of learning-based techniques for delivery optimization, including reinforcement learning, graph neural networks, and other data-driven routing models~\cite{de2021end, gao2021deep, li2018learning}. 
These approaches learn adaptive policies from historical or simulated demand and traffic data, enabling flexibility under dynamic conditions.

\noindent \textbf{Reinforcement Learning-Based Approaches:} 
RL has been applied to logistics through multi-agent fleet control models \cite{fleetmanagement, haliem2021distributed, bi2024truck} and freight routing frameworks such as DeepFreight \cite{deepfreight}. While effective in structured or low-dimensional settings, these approaches often rely on centralized critics or simplified problem assumptions, limiting scalability for real-time OFD. DeliverAI \cite{mehra2024deliverai,10705350} introduced multi-hop, path-sharing delivery via MARL but depends on a centralized architecture and a fully connected graph, preventing deployment at city scale.

We implement a MADDPG baseline \cite{lowe2017multi} for comparison. Although CTDE enables decentralized execution, training remains expensive, scales quadratically with fleet size, and supports only pairwise bundling interactions. Moreover, policies must be retrained whenever demand or travel conditions shift. In contrast, UMST requires no training, adapts instantly to environmental changes, and still matches or exceeds MADDPG performance—even though MADDPG operates over the full graph while UMST is restricted to a reduced backbone.

\noindent \textbf{Graph Neural Network-Based Approaches:} 
GNNs have been explored for routing and delivery optimization through spatial–temporal message passing \cite{wen2022graph2route}, autonomous vehicle navigation \cite{kazmi2024groutenet}, and driver behavior modeling \cite{aldhahri2025gnnrm}. These methods leverage historical data to learn rich structural patterns, often achieving strong predictive accuracy.

However, GNN-based routing inherits key limitations of learning-driven approaches: models require extensive training data, incur high computational cost, and generalize poorly across cities or changing demand distributions. They also operate as black-box systems without guarantees on connectivity or robustness to edge failures. UMST avoids these issues entirely: it requires no training, provides interpretable connectivity guarantees, and achieves performance competitive with GNN baselines despite using only a sparse, redundancy-enhanced backbone.

\section{UMST-Based Delivery Backbone}
\label{sec:methodology}

This section describes the systematic construction of the UMST network and its integration into our delivery optimization framework. The methodology consists of four major phases: data acquisition and preprocessing, spatial decomposition and hotspot identification, graph construction, and UMST generation.

\begin{figure*}[t]
\centering
\includegraphics[width=0.67\textwidth]{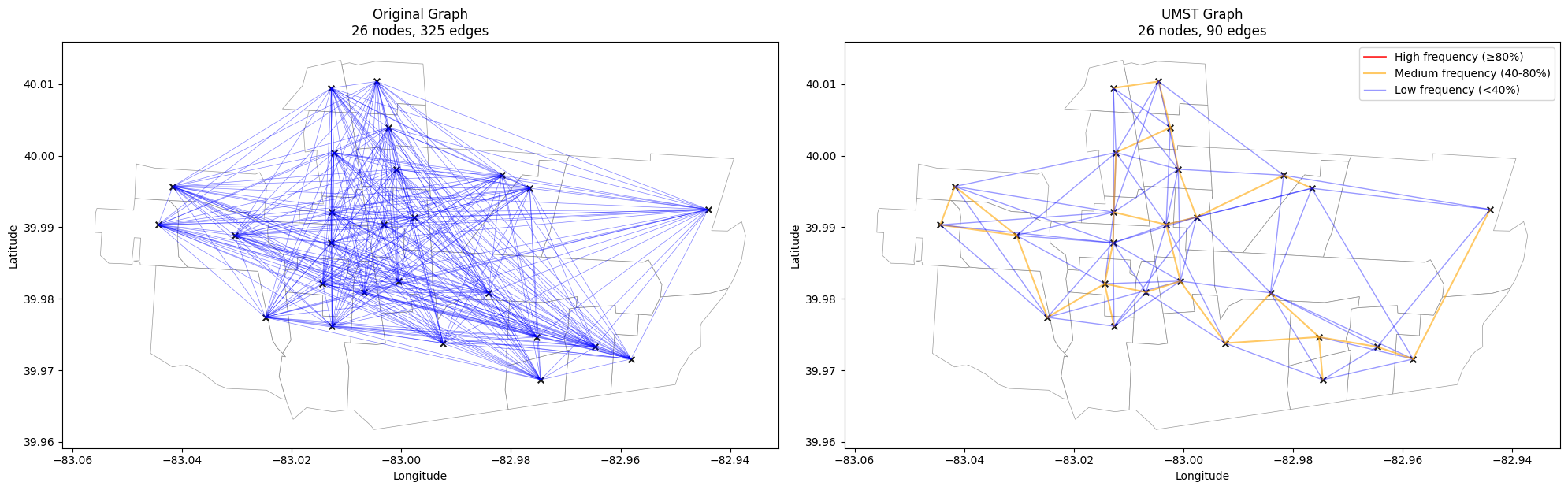}
\includegraphics[width=0.31\textwidth]{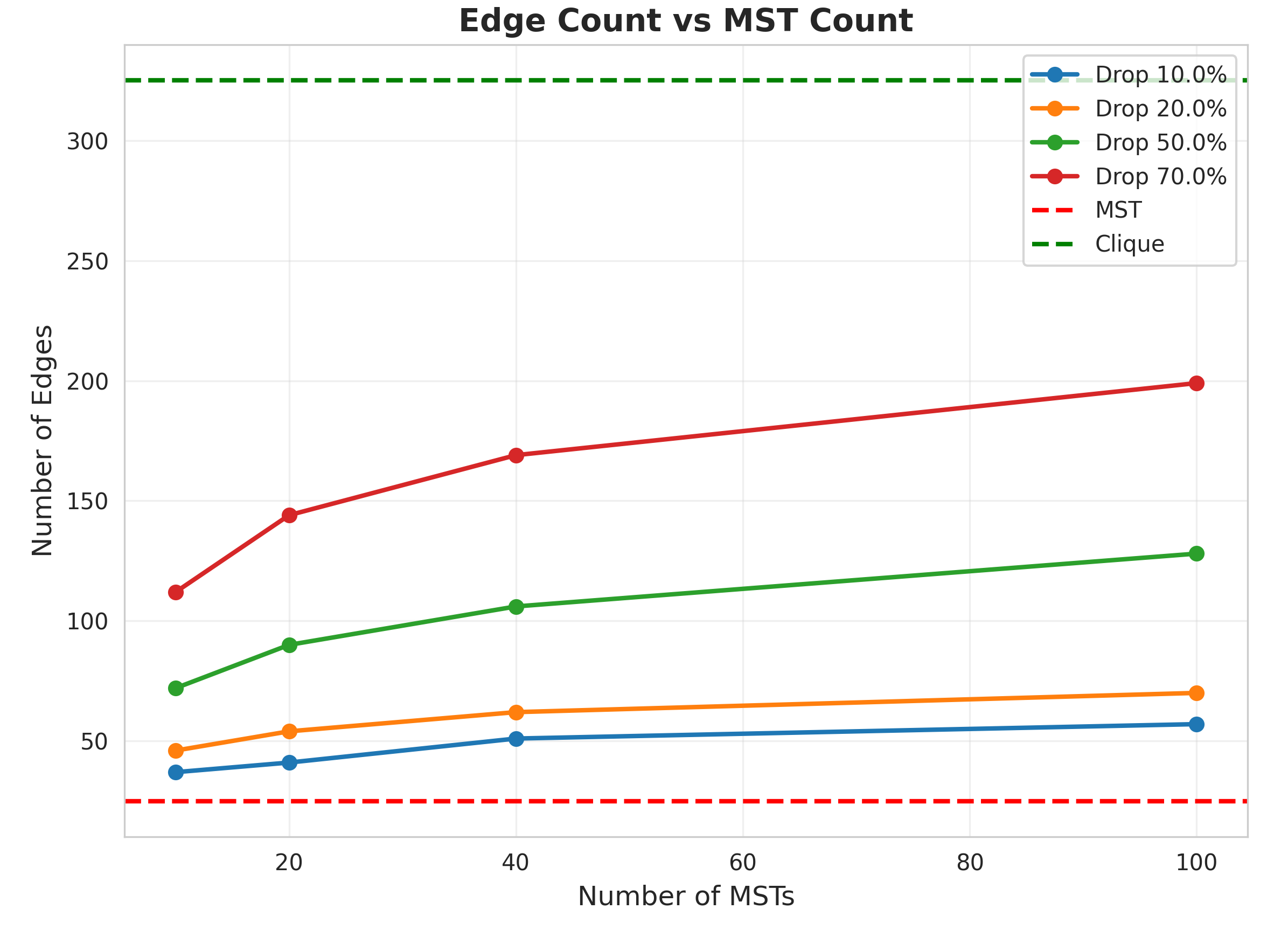}
\caption{Construction of UMST Backbone for Columbus (downtown), Ohio. From left to right: (a) completely connected hotspot graph, (b) corresponding UMST graph, and (c) UMST edge count variations.}
\label{fig:UMST formation}
\end{figure*}

\subsection{Data Acquisition and Preprocessing}
\label{subsec:data_acquisition}

Our network construction begins with three urban data sources. OpenStreetMap provides the basic geography of each city, including road networks and the locations of restaurants, distribution centers, and residential buildings. Census tract boundaries are used to divide the city into meaningful spatial units with consistent population characteristics. Finally, we use GraphHopper to retrieve real-world distances and travel times along the road network, ensuring that all routing decisions reflect the actual structure of the streets rather than simple Euclidean geometry.

\subsection{Spatial Decomposition and Hotspot Identification}
\label{subsec:spatial_decomposition}

The city is decomposed into census tracts, and each tract is represented by a single hotspot. We collect all producer and consumer locations inside a tract and summarize them by computing the centroid of that region. This centroid becomes the hotspot node associated with that tract and serves as the representative point for both demand and supply. The set of all hotspots forms the node set of our delivery network.

\subsection{Graph Construction}
\label{subsec:graph_construction}

We begin by constructing a complete weighted graph over all hotspots. Every pair of hotspots is connected by a conceptual edge, representing the possibility of travel between those locations. For each such pair, we query GraphHopper to obtain the actual road-network shortest-path distance and travel time. These values replace purely geometric distances and ensure that all subsequent computations use realistic transportation costs derived from the underlying city road network.

The resulting fully connected, road-weighted graph serves as the input to the UMST construction procedure. Unlike approaches that prune long-range edges or enforce local neighborhood restrictions, we retain the full connectivity structure so that the UMST can explore diverse combinations of edges during randomized MST sampling. This preserves global structure, enables richer redundancy patterns, and provides a basis for generating the sparse but well-connected delivery backbone used for routing and bundling in later stages.

\subsection{Union of Minimum Spanning Trees (UMST)}
\label{subsec:umst_generation}


Urban delivery network design must balance two fundamentally competing priorities: keeping routing costs low while maintaining high reliability under dynamic conditions. Extremely sparse structures - such as the classical Minimum Spanning Tree (MST) - do not optimize total travel distance, offer almost no flexibility, making the system vulnerable to congestion, local disruptions, but are extremely efficient to compute and offer several opportunities for bundling. At the opposite extreme, a fully connected graph provides excellent redundancy, gives the shortest possible distances, but is far too dense and operationally expensive to be practical.

In this paper, we introduce the Union Minimum Spanning Tree (UMST) framework, which is designed to operate precisely between these two extremes. Rather than committing to a single, rigid MST, we generate multiple MSTs by repeatedly introducing small random perturbations to the underlying graph. Each perturbation removes a small fraction of edges, forcing the MST algorithm to select alternative low-cost connections that a single tree would never reveal. The union of these trees forms a lightweight but significantly richer routing backbone that preserves the cost efficiency of MST-like structures while introducing enough diversity to support reliable planning.

This union graph expands the feasible routing space without drifting toward the combinatorial explosion of a dense graph. The resulting topology enables more flexibility in avoiding local bottlenecks, provides alternate paths when disruptions occur, and increases the opportunities for bundling multiple orders along similar routes. When we normalize and jointly evaluate cost- and reliability-related metrics, the UMST consistently achieves the strongest combined performance across the design space. It avoids the fragility of ultra-sparse networks and the overhead of overly dense ones, emerging as the most balanced and mutually beneficial compromise for urban delivery routing.



Using downtown Columbus, Ohio, as an example, we begin with 26 census tracts and extract producer/consumer activity from OpenStreetMap. A hotspot is placed at the geometric centroid of each tract, yielding 26 nodes whose spatial distribution captures aggregated supply and demand (see figure in supplemental file). We then construct the complete hotspot graph by linking every pair of hotspots, resulting in 325 edges (see Fig.~\ref{fig:UMST formation}(a)). To introduce structural diversity, we generate multiple perturbed versions of this graph by randomly dropping 50\% of its edges and computing a minimum spanning tree (MST) on each. The UMST is formed by taking the union of 20 such MSTs, producing a 90-edge sparse graph that preserves full connectivity while reducing the original graph by approximately 72\%. Edges that appear frequently across the MST ensemble form the high-frequency backbone that guides routing and opportunistic bundling. 

This frequency-weighted union preserves the essential connectivity patterns of the city while dramatically reducing graph complexity. High-frequency edges ($\geq 80\%$ occurrence) form the backbone, medium-frequency edges (40-80\%) provide secondary alternatives, and low-frequency edges ($<40\%$) act as fallback links. The resulting topology is both sparse and robust: while the UMST is only 28\% as dense as the full graph, its structure reflects consistent connectivity patterns across the MST ensemble. 

Additional experiments with different drop rates and number of MSTs (see Fig.~\ref{fig:UMST formation}(c)) show how the number of UMST edges compares with the completely connected hotspot graph ("clique"). Also, experiments confirm that (see supplemental file for the \textbf{figure}) certain key edges persist regardless of the weighting scheme (distance, travel time, or reliability), demonstrating that the UMST captures stable, city-wide routing structure while maintaining multiple viable fallback paths for resilience.

\subsubsection{UMST Construction Algorithm}

Let $G = (H, E)$ be the filtered hotspot graph with edge weights given by geographical distances. To construct the UMST, we generate $K$ perturbed MSTs using randomized edge deletion. In each iteration we remove a small proportion $\rho$ of edges, compute an MST on the remaining graph, and accumulate all edges that ever appear. The full procedure is provided in Algorithm~\ref{alg:umst_construction}.

\begin{algorithm}[H]
\caption{UMST Construction via Randomized Edge Dropping}
\label{alg:umst_construction}
\begin{algorithmic}[1]
\STATE \textbf{Input:} Graph $G = (H, E)$, number of MSTs $K$, drop rate $\rho = 0.1$
\STATE \textbf{Output:} UMST graph $G_{\text{UMST}} = (H, E_{\text{UMST}})$
\STATE Initialize $E_{\text{UMST}} \leftarrow \emptyset$
\FOR{$k = 1$ to $K$}
    \STATE $E_k \leftarrow E$
    \STATE $n_{\text{drop}} \leftarrow \lfloor \rho \cdot |E| \rfloor$
    \STATE Randomly remove $n_{\text{drop}}$ edges from $E_k$
    \STATE Compute MST $T_k$ on $(H, E_k)$
    \STATE $E_{\text{UMST}} \leftarrow E_{\text{UMST}} \cup \text{edges}(T_k)$
\ENDFOR
\STATE \textbf{return} $G_{\text{UMST}} = (H, E_{\text{UMST}})$
\end{algorithmic}
\end{algorithm}

In our experiments we use different values for the hypeparameters $K \in \{10, 20, 40, 100\}$ MSTs with $\rho \in \{0.1, 0.2, 0.5, 0.7\}$. Each iteration exposes a distinct low-cost structure, and the resulting union includes all edges that appear in at least one MST. 


\textbf{Theoretical Basis for using UMST:} Two results form the basis for using UMST - (i) For large enough sampling of trees (specifically $\mathcal{O}(\log(|E|))$) we can get good approximations to the dense graph and (ii) the ``stretch'' (error in approximating shortest distances) reduces \emph{exponentially} as $\exp(-c K)$ where $K$ is the number of samples of MSTs considered. Both these results are justified in the supplements.

\section{Problem Definition}
\label{sec:problem-definition}

We model the urban delivery environment using a UMST-derived graph $G' = (H, E')$
, where $H$ is the set of hotspots and $E'$ is the set of UMST edges between hotspots. Each edge 
$e=(u,v)\in E'$ has an associated travel time $\tau_e>0$. The UMST provides a sparse, low-diameter backbone that defines a unique next-hop direction for routing between any pair of hotspots.

\subsection{Vehicles and Requests}

Let $\mathcal{V}$ be the fleet of vehicles. Each vehicle $k \in \mathcal{V}$ has a capacity limit $C(k)$ and follows routes composed exclusively of edges in $E'$. Each delivery request is
$r = (p_r, d_r, \tau_p, \tau_d),$
where $p_r$ and $d_r$ are pickup and drop-off hotspots, $\tau_p$ is the earliest pickup time, and $\tau_d$ is the delivery deadline. A request is successful if it is picked up no earlier than $\tau_p$ and delivered by $\tau_d$.

For each request $r$, we denote by
$\pi_r = (h_0, h_1, \ldots, h_L)$
the UMST path from $p_r$ to $d_r$, and by $\mathrm{next}(h)$ the UMST next-hop from hotspot $h$ along this path.

\subsection{Opportunistic Bundling and Bundle Merging}

Routing occurs along the UMST backbone using a greedy, local dispatching mechanism at each hotspot. At hotspot $h$, all active requests whose UMST next hop equals $v = \mathrm{next}(h)$ form a candidate bundle:
\[
B(h,v) = \{ r : p_r = h,\; \mathrm{next}(h) = v \}.
\]

A vehicle departing from $h$ toward $v$ may carry any subset of $B(h,v)$ such that its capacity is not exceeded. This induces an \emph{opportunistic bundling} mechanism: requests sharing a UMST prefix naturally travel together without requiring detours or global optimization.

The model also supports \emph{bundle merging}. If a vehicle arrives at hotspot $h$ carrying a bundle $B_{\mathrm{in}}$ and additional local requests at $h$ share the same next-hop direction $v$, they may merge: $B_{\mathrm{out}} = B_{\mathrm{in}} \cup B(h,v)$ and $|B_{\mathrm{out}}| \le C(k).$ Thus, both individual requests and existing bundles can combine into larger bundles whenever their UMST paths align.

\subsection{Vehicle Routes}


Each vehicle $k$ executes a route $R_k = (h_0, h_1, \ldots, h_m)$, where $(h_j, h_{j+1}) \in E'$ for all $j$. Let $t_k(h)$ denote the vehicle’s arrival time at hotspot $h$. A delivery request $r$ carried by vehicle $k$ is feasible if \(t_k(p_r) \ge \tau_p \text{ and } t_k(d_r) \le \tau_d\). The total travel time of vehicle $k$ is \(T(R_k) = \sum_{j=0}^{m-1} \tau_{(h_j,h_{j+1})}\).

\subsection{Feasibility Constraints}

We next formalize the constraints that govern request assignments and vehicle operations.


\paragraph{Assignment.}
Each accepted request satisfies $\sum_{k \in \mathcal{V}} x_{rk} = 1$ $\forall r$.

\paragraph{Precedence.}
For any request assigned to vehicle $k$, the pickup precedes the drop-off along the route:
$p_r \prec d_r $ in  $R_k \ \forall r,k \text{ with } x_{rk}=1.$

\paragraph{Capacity.}
The load of any vehicle never exceeds its capacity:
$\mathrm{Load}_k(\ell) \le C(k) \ \forall k,\ell.$

\paragraph{UMST Feasible Edges.}
Vehicles may traverse only edges in the UMST graph:
$R_k \subseteq E' \ \forall k.$


\paragraph{Travel-Time Computation.} The fleet-wide travel time is \(T = \sum_{k \in \mathcal{V}} T(R_k)\).

\subsection{Objective}


Let $S$ denote the set of successfully delivered requests. The goal is to select feasible UMST-based routes and bundle assignments that maximize $|S|$ while keeping $T$ low.

In the baseline setting, routing follows shortest paths on the original road network $G$. In contrast, our framework operates on the UMST graph $G'$, whose structured redundancy enables opportunistic bundling, scalable merging of bundles, and more reliable delivery performance while retaining sparsity and computational efficiency.

\section{Performance Analysis and Results}
\label{sec:umst-results}

This section presents the experimental results for the UMST approach,
followed by a description of our simulation environment and the
evaluation metrics.

\subsection{Simulation Setup and Workload Generation}
\label{subsec:simulation_setup}

Our experiments use real-world data from Columbus (26 census tracts)
and Chicago (31 census tracts), focusing on core urban areas.
Each simulation run spans one hour and generates ~9234 delivery
requests. Requests arrive according to a two-peak Gaussian mixture
with peaks at 0.25 and 0.75 of the window (approximately 15 and 45
minutes) and temporal spread $\sigma=10$. Every delivery is limited
to a maximum trip duration of 1{,}800 seconds and uses buffer times
from the \texttt{rangewise} strategy. The arrival intensity is
modeled as a sum of Gaussians,
\[
  y(t)=\sum_{i=1}^{k}\frac{1}{\sigma\sqrt{2\pi}}\exp\!\left(-\tfrac{1}{2}\left(\tfrac{t-\mu_i}{\sigma}\right)^{2}\right),
\]
with $\mu_i$ set to the peak locations; $y(t)$ is then scaled to
produce the minute-level request volumes used by the simulator. This
compact, reproducible workload produces realistic temporal
variability while keeping experimental conditions consistent across
UMST and baseline evaluations.
Values for various hyperparameters for the simulation were finalized
empirically (see Table of hyperparameters and delivery request
distribution graph in the supplemental file).

\subsection{Baseline Comparison}

We compare UMST against two learning-based baselines: a GNN routing
model~\cite{wen2022graph2route} and the MADDPG
framework~\cite{lowe2017multi}. Both require offline training and
retraining under changing conditions. We also compare UMST with two
structural baselines - a completely connected hotspot graph (aka.
Clique) and the MST.



\subsection{Evaluation Metrics}

We report five primary metrics that characterize system reliability,
delivery efficiency, and bundling behavior.

\paragraph{Success Rate.}
Success rate is the fraction of requests successfully delivered. If
$N_{\text{succ}}$ is the number of completed deliveries and
$N_{\text{req}}$ is the total number of requests, then the metric is
defined as $\text{SuccessRate} = N_{\text{succ}} / N_{\text{req}}$.
This captures overall service feasibility. Higher values are better.

\paragraph{Average (Completion) Time.}
For each successful delivery $i$ with completion time $t_i$, the
average completion time is given by $\text{AvgTime} = (1 /
N_{\text{succ}}) \sum_i t_i$. Lower values indicate faster delivery
performance and more efficient routing. Lower values are better.

\paragraph{Total Vehicle Distance (KM).}
Let $d_v$ denote the total distance traveled by all the vehicles
($v$). The system-level travel cost is computed as
$\text{VehicleDistance} = \sum_{v \in \mathcal{V}} d_v$. This metric
reflects the routing efficiency induced by the network structure.
Lower values are better.

\paragraph{Total Package Distance (KM).}
Let $d_p$ denote the total distance traveled by all the delivery
packets ($p$). The system-level travel cost is computed as
$\text{PackageDistance} = \sum_{p \in \mathcal{P}} d_p$. This metric
reflects the bundling efficiency induced by the UMST model. Higher
value of $d_p$ compared to $d_v$  signifies the distance saved
through bundling. Lower values are better.

\paragraph{Bundling Participation.}
Bundling participation is defined as the total number of deliveries
that shared a vehicle with one or more other deliveries at any point
during service. It represents how many deliveries took part in
bundling. Higher values are better.

\subsection{Simulation Results}

We evaluate UMST in three stages: first by establishing its
structural position as a balanced trade-off between MST and a
complete hotspot graph, then by quantifying gains enabled by order
bundling, and finally by comparing its delivery performance against
graph- and learning-based baselines across five key metrics.
Together, these results show that UMST provides both structural
efficiency and practical performance under dynamic demand. All
methods are evaluated under identical demand.

\subsubsection{Structural Comparison Results}
The MST represents the most cost-efficient but brittle backbone,
while the complete graph offers maximal routing flexibility at a
prohibitive complexity. Our results show that UMST consistently
occupies a middle-ground trade-off zone, retaining much of the
efficiency of an MST while introducing enough path diversity to
approach the robustness of denser graphs.

\begin{figure}[ht]
  \centering
  \includegraphics[width=0.85\columnwidth]{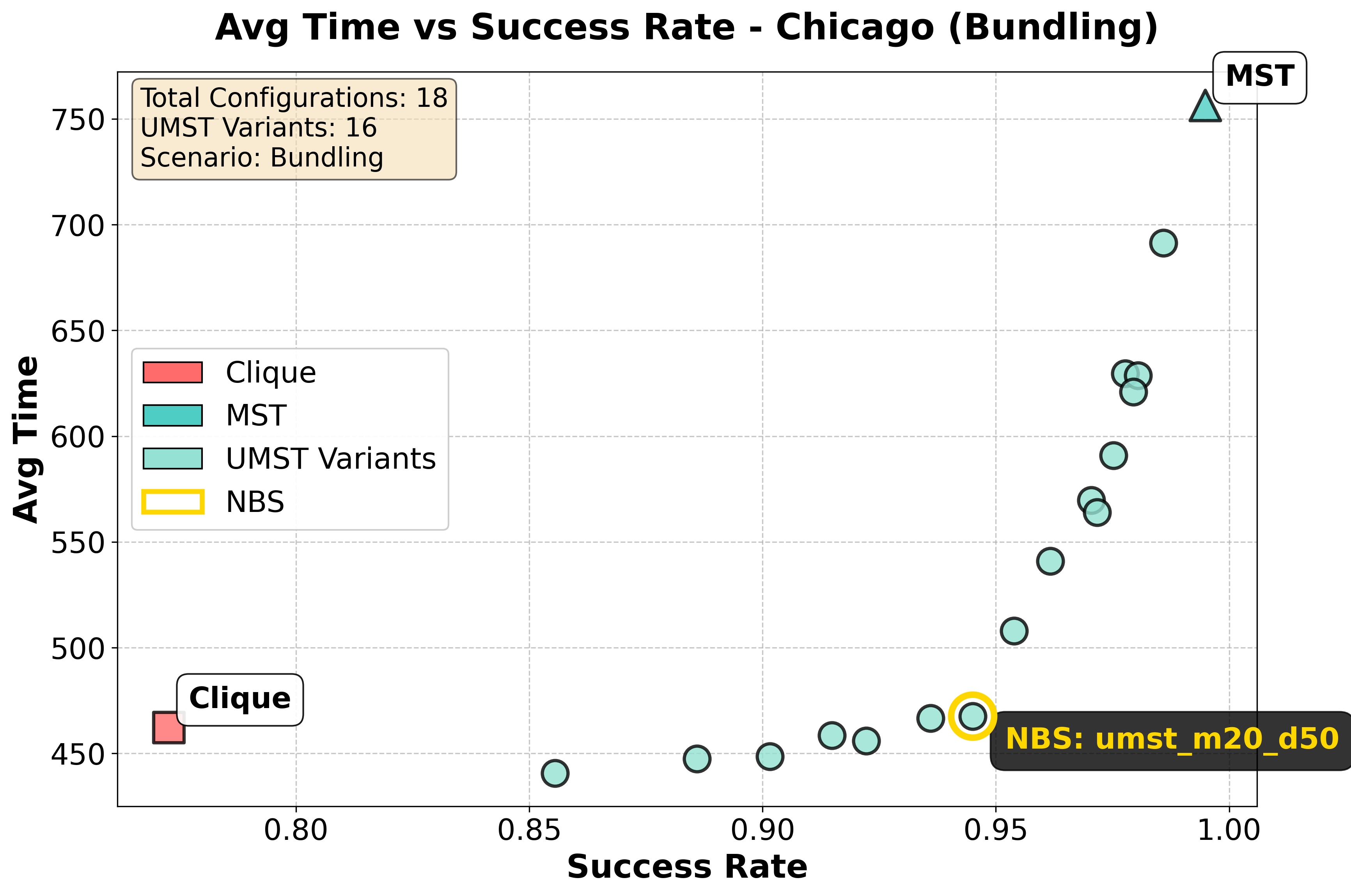}
  \caption{Average completion time vs.\ success rate in Chicago under
    bundling. UMST variants dominate the trade-off curve, with the Nash
    Bargaining Solution (NBS) representing the best balance between
  speed and reliability.}
  \label{fig:chicago_bundling_tradeoff_time}
\end{figure}

Figure~\ref{fig:chicago_bundling_tradeoff_time} shows the
success–time trade-off for all configurations in the Chicago bundling
scenario. The MST baseline achieves the highest success rate, close
to 0.99, but also incurs the longest completion time at over 800
units. In contrast, the Clique (completely connected hotspot graph)
baseline offers low completion time around 440 units but much lower
reliability near 0.78. UMST configurations lie between these extremes
and form a clear trade-off curve. Among them,
$\texttt{umst\_m20\_d50}$ provides the best balance, achieving
roughly 0.92 success while keeping completion time near 480 units.
This places it closest to the desirable region of high reliability
and low delay, confirming that UMST effectively balances competing objectives.

\begin{figure}[ht]
  \centering
  \includegraphics[width=0.85\columnwidth]{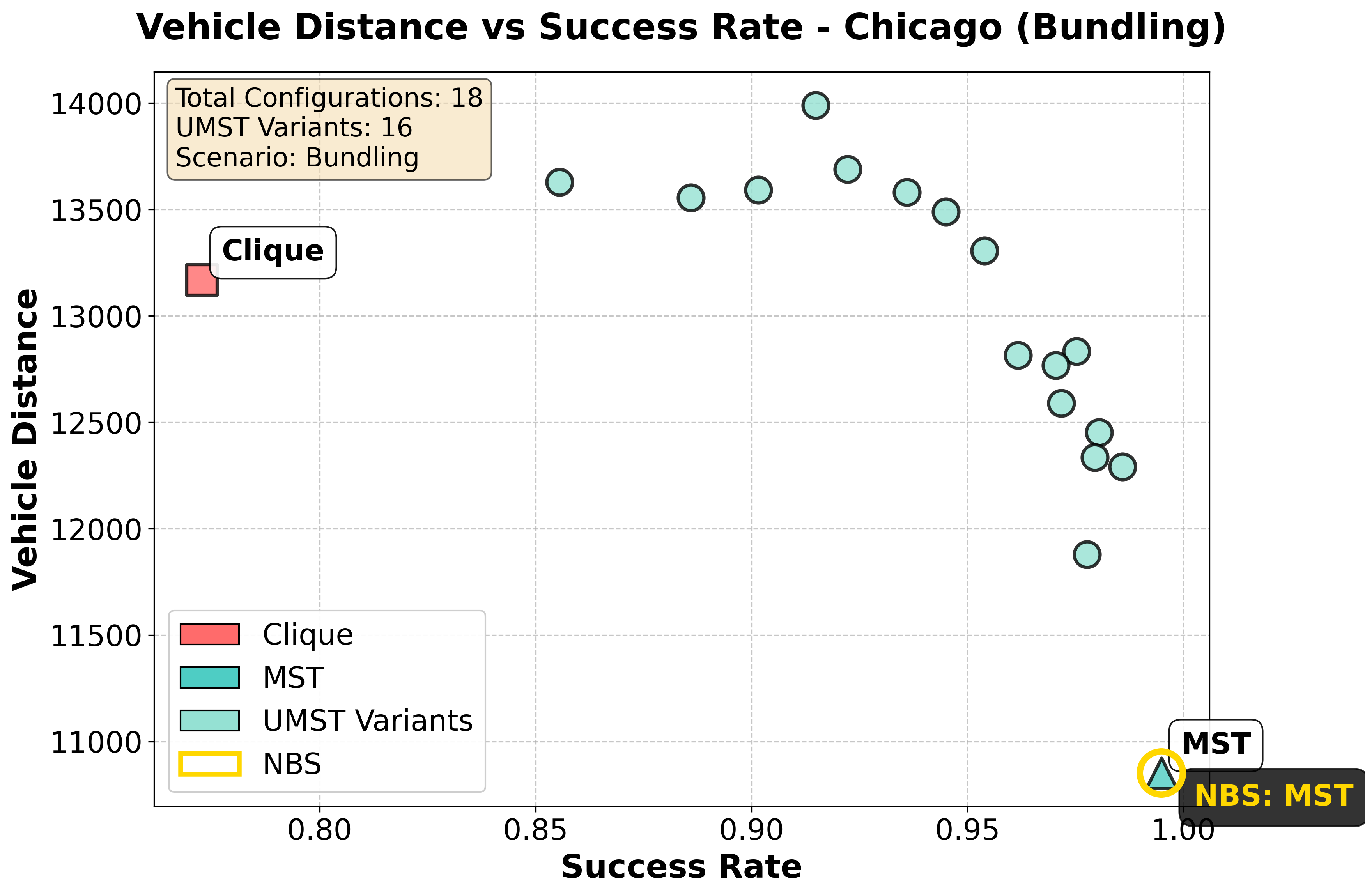}
  \caption{Vehicle distance vs.\ success rate in Chicago under
    bundling. UMST variants dominate the trade-off curve. The Nash
    Bargaining Solution (NBS) represents the best balance between
  efficiency and reliability.}
  \label{fig:chicago_bundling_tradeoff_dist}
\end{figure}

Figure~\ref{fig:chicago_bundling_tradeoff_dist} presents the
vehicle-distance versus success-rate trade-off. MST attains the
lowest distance and highest success rate but lacks robustness to
disruptions. The Clique baseline performs poorly on both metrics,
with high travel distance and low success. UMST again occupies the
middle ground, with the strongest configurations achieving success
rates above 0.90 at moderate distance overheads of about 13–16\%
relative to MST. These UMST variants significantly outperform Clique
while maintaining resilience MST cannot provide. Overall, UMST
achieves near-MST reliability with manageable distance increases,
demonstrating a practical balance between efficiency, reliability,
and robustness.

\begin{table*}[h]
  \centering
  \caption{Performance comparison (Mean ± Std, 10 Runs) of Clique,
    UMST, and MST backbones across Chicago and Columbus (9,234
    deliveries per scenario). Bold values indicate better performance
    in each comparison. Arrows indicate optimization direction:
    $\uparrow$ higher is better, $\downarrow$ lower is better. UMST
    consistently occupies the middle ground between dense and sparse
    extremes, achieving substantially higher success rates than Clique
    while avoiding the excessive delivery times of MST. Across both
    cities, UMST maintains competitive vehicle distance and strong
    order bundling participation, demonstrating its ability to balance
    delivery reliability, efficiency, and operational flexibility more
  effectively than either baseline. }
  \label{tab:simulation_results}
  \small
  \begin{tabular}{llcccccc}
    \toprule
    \textbf{City} & \textbf{Graph} & \textbf{Success} & \textbf{Avg
    Time} & \textbf{Vehicle} & \textbf{Package} & \textbf{Saved} &
    \textbf{Bundling} \\
    & \textbf{Type} & \textbf{Rate (\%)} $\uparrow$ & \textbf{(sec)}
    $\downarrow$ & \textbf{Dist. (km)} $\downarrow$ & \textbf{Dist.
    (km)} $\downarrow$ & \textbf{Dist. (km)} $\uparrow$ &
    \textbf{Participation} $\uparrow$ \\
    \midrule

    \multirow{4}{*}{\rotatebox{90}{\textbf{Chicago}}}
    & Clique & $73.82 \pm 0.24$ & $\mathbf{477.63 \pm 2.19}$ &
    $\mathbf{13{,}134.44 \pm 119.76}$ & $\mathbf{17{,}771.19 \pm
    152.93}$ & $6{,}045.61$ & $4{,}520 \pm 24$ \\
    & UMST & $\mathbf{88.07 \pm 0.07}$ & $500.37 \pm 1.02$ &
    $13{,}480.29 \pm 63.43$ & $22{,}952.53 \pm 54.27$ &
    $\mathbf{10{,}427.44}$ & $\mathbf{7{,}690 \pm 36}$ \\
    \cmidrule(lr){2-8}
    & MST & $\mathbf{98.78 \pm 0.10}$ & $810.03 \pm 5.50$ &
    $\mathbf{10{,}853.36 \pm 45.29}$ & $34{,}038.78 \pm 248.44$ &
    $\mathbf{23{,}400.54}$ & $\mathbf{9{,}092 \pm 10}$ \\
    & UMST & $88.07 \pm 0.07$ & $\mathbf{500.37 \pm 1.02}$ &
    $13{,}480.29 \pm 63.43$ & $\mathbf{22{,}952.53 \pm 54.27}$ &
    $10{,}427.44$ & $7{,}690 \pm 36$ \\
    \midrule

    \multirow{4}{*}{\rotatebox{90}{\textbf{Columbus}}}
    & Clique & $82.07 \pm 0.23$ & $\mathbf{599.18 \pm 2.00}$ &
    $31{,}354.99 \pm 91.47$ & $\mathbf{42{,}992.83 \pm 122.80}$ &
    $\mathbf{2{,}058.15}$ & $4{,}833 \pm 47$ \\
    & UMST & $\mathbf{96.05 \pm 0.12}$ & $642.72 \pm 1.76$ &
    $\mathbf{26{,}826.79 \pm 86.83}$ & $48{,}369.45 \pm 207.84$ &
    $689.18$ & $\mathbf{7{,}670 \pm 30}$ \\
    \cmidrule(lr){2-8}
    & MST & $\mathbf{99.55 \pm 0.02}$ & $944.78 \pm 8.00$ &
    $\mathbf{20{,}236.58 \pm 117.18}$ & $61{,}913.38 \pm 556.74$ &
    $396.18$ & $\mathbf{9{,}018 \pm 16}$ \\
    & UMST & $96.05 \pm 0.12$ & $\mathbf{642.72 \pm 1.76}$ &
    $26{,}826.79 \pm 86.83$ & $\mathbf{48{,}369.45 \pm 207.84}$ &
    $\mathbf{689.18}$ & $7{,}670 \pm 30$ \\
    \bottomrule
  \end{tabular}
\end{table*}

\begin{table*}[h]
\centering
\caption{
Impact of order bundling on UMST performance across Chicago and Columbus (Mean ± Std, 10 Runs). Bold values indicate better performance in each comparison. Arrows indicate optimization direction: $\uparrow$ higher is better, $\downarrow$ lower is better. Negative delay values indicate early arrivals. Bundling introduces only modest reductions in service quality while delivering dramatic operational gains, cutting total vehicle distance by 44–53\% and achieving high consolidation rates. The results show that UMST effectively converts limited increases in delivery time into substantial efficiency, cost, and emission savings, confirming bundling as a highly favorable trade-off when supported by a flexible backbone.}
\label{tab:bundling_no_bundling_cities}
\small
\begin{tabular}{lcccc}
\toprule
 & \multicolumn{2}{c}{\textbf{Chicago}} & \multicolumn{2}{c}{\textbf{Columbus}} \\
\cmidrule(lr){2-3} \cmidrule(lr){4-5}
\textbf{Metric} & \textbf{No Bundling} & \textbf{Bundling} & \textbf{No Bundling} & \textbf{Bundling} \\
\midrule
Total Deliveries & $9{,}234$ & $9{,}234$ & $11{,}550$ & $11{,}550$ \\
Completion Rate (\%) $\uparrow$ & $\mathbf{100.00 \pm 0.00}$ & $98.65 \pm 0.04$ & $\mathbf{97.72 \pm 0.15}$ & $96.75 \pm 0.16$ \\
Success Rate (\%) $\uparrow$ & $\mathbf{100.00 \pm 0.00}$ & $96.08 \pm 0.07$ & $\mathbf{97.72 \pm 0.15}$ & $92.53 \pm 0.23$ \\
\midrule
Avg Time (sec) $\downarrow$ & $\mathbf{591.97 \pm 3.23}$ & $641.13 \pm 3.34$ & $\mathbf{578.43 \pm 1.98}$ & $653.43 \pm 2.08$ \\
Avg Delay (sec) $\downarrow$ & $\mathbf{-362.99 \pm 1.02}$ & $-312.10 \pm 0.93$ & $\mathbf{-251.36 \pm 0.59}$ & $-174.90 \pm 0.90$ \\
\midrule
Vehicle Distance (km) $\downarrow$ & $48{,}941.87 \pm 276.44$ & $\mathbf{27{,}146.64 \pm 148.14}$ & $27{,}491.72 \pm 107.52$ & $\mathbf{12{,}917.15 \pm 43.13}$ \\
Package Distance (km) $\downarrow$ & $48{,}941.87 \pm 276.44$ & $\mathbf{48{,}226.63 \pm 266.86}$ & $27{,}491.72 \pm 107.52$ & $\mathbf{27{,}164.90 \pm 110.39}$ \\
Distance Saved (km) $\uparrow$ & $0.00$ & $\mathbf{21{,}795.23}$ & $0.00$ & $\mathbf{14{,}574.56}$ \\
\bottomrule
\end{tabular}

\end{table*}

\subsubsection{Comparison of Graph-Based Bundling Strategies}
Table~\ref{tab:simulation_results} compares three graph structures (Clique, UMST, and MST) across Chicago and Columbus with 9,234
deliveries per scenario.

\noindent\textbf{Success Rate and Efficiency Trade-offs.} UMST demonstrates a balanced position between the extremes. In Chicago, UMST achieves 88.07\% ($\pm$0.07) success rate with 500.37 sec average time, positioned between Clique's lower success (73.82\%) and MST's higher success (98.78\%) but much longer time (810.03 sec). Columbus shows similar patterns where UMST achieves 96.05\% success with 642.72 sec delivery time, significantly outperforming Clique (82.07\%) while maintaining faster service than MST (944.78 sec).

\noindent\textbf{Operational Efficiency.} UMST's vehicle distance remains competitive (13,480.29 km in Chicago, 26,826.79 km in Columbus) while enabling substantial bundling. Chicago shows 7,690 ($\pm$36) participating deliveries saving 10,427.44 km, while Columbus achieves 7,670 ($\pm$30) participation. MST achieves the highest distance savings (23,400.54 km Chicago, though minimal 396.18 km Columbus) but at the cost of significantly slower deliveries. The results confirm that UMST provides superior balance by avoiding Clique's poor reliability and MST's excessive delays while maintaining strong bundling efficiency.

\subsubsection{Bundling vs.\ No-Bundling Comparison}
Table~\ref{tab:bundling_no_bundling_cities} evaluates UMST with and without bundling across Chicago (9,234 deliveries) and Columbus (11,550 deliveries), averaged over 10 runs.

\noindent\textbf{Service Quality Trade-offs.} Bundling introduces modest reductions in service quality. Success rates decrease from 100.00\% to 96.08\% in Chicago and from 97.72\% to 92.53\% in Columbus. Average delivery times increase moderately by 8.3\% in Chicago (591.97 s to 641.13 s) and 13.0\% in Columbus (578.43 s to 653.43 s). Negative average delays ($-312.10$ s and $-174.90$ s) confirm that deliveries remain well ahead of deadlines despite consolidation delays.

\noindent\textbf{Operational Efficiency Gains.} Bundling produces substantial resource savings. Vehicle distance drops 44.5\% in Chicago (48,941.87 km to 27,146.64 km) and 53.0\% in Columbus (27,491.72 km to 12,917.15 km), saving 21,795.23 km and 14,574.56 km respectively. High bundling rates of 82.89\% in Chicago (7,654 deliveries in 7,555 bundles) and 74.65\% in Columbus (8,621 deliveries in 6,909 bundles) demonstrate effective consolidation with low variance across runs.

\noindent\textbf{Cost-Benefit Assessment.} A 3 to 8\% reduction in success rate yields 44 to 53\% lower vehicle distance. For an operation with 10,000 daily deliveries, this corresponds to avoiding 20,000 to 25,000 km of travel, producing daily savings of \$10,000 to 25,000 in fuel and maintenance. Annualized, these efficiencies reach millions of dollars while reducing emissions proportionally. The modest increase in delivery time remains within common service-level thresholds, confirming a favorable efficiency-quality trade-off.

\begin{table*}[t]
  \centering
  \small
  \caption{Comparison of UMST with learning-based baselines (MADDPG
    and GNN over UMST backbone) across Columbus and Chicago (9,234
    deliveries). Bold values indicate better performance in each
    comparison. Arrows indicate optimization direction: $\uparrow$
    higher is better, $\downarrow$ lower is better. Negative delay
    values indicate early arrivals. UMST achieves strong success rates
    and competitive distance savings while significantly outperforming
    learned methods in delivery time, offering a robust, training-free
    alternative that balances reliability, efficiency, and practical
  deployability.}
  \label{tab:comparison_cities}
  \setlength{\tabcolsep}{4pt}
  \begin{tabular}{l|cc|cc||cc|cc}
    \toprule
    & \multicolumn{4}{c||}{Columbus} & \multicolumn{4}{c}{Chicago} \\
    \cmidrule(lr){2-5} \cmidrule(lr){6-9}
    Metric & MADDPG & UMST & GNN & UMST & MADDPG & UMST & GNN & UMST \\
    \midrule
    Total Deliveries & 9,234 & 9,234 & 9,234 & 9,234 & 9,234 & 9,234
    & 9,234 & 9,234 \\
    Completed Del. $\uparrow$ & $\mathbf{9,234}$ & 9,103 &
    $\mathbf{9,234}$ & 9,103 & $\mathbf{9,234}$ & 9,080 &
    $\mathbf{9,234}$ & 9,080 \\
    Successful Del. $\uparrow$ & 8,472 & $\mathbf{8,868}$ &
    $\mathbf{9,234}$ & 8,868 & 8,740 & $\mathbf{8,720}$ &
    $\mathbf{9,234}$ & 8,720 \\
    Failed Del. $\downarrow$ & 762 & $\mathbf{235}$ & $\mathbf{0}$ &
    235 & 494 & $\mathbf{360}$ & $\mathbf{0}$ & 360 \\
    Success Rate $\uparrow$ & 0.918 & $\mathbf{0.960}$ &
    $\mathbf{1.0}$ & 0.960 & 0.947 & $\mathbf{0.944}$ &
    $\mathbf{1.0}$ & 0.944 \\
    Completion Rate $\uparrow$ & $\mathbf{1.0}$ & 0.986 &
    $\mathbf{1.0}$ & 0.986 & $\mathbf{1.0}$ & 0.983 & $\mathbf{1.0}$ & 0.983 \\
    Training Time (min) $\downarrow$ & $293.09$ & $\mathbf{0}$ &
    $29.41$ & $\mathbf{0}$ & $276.63$ & $\mathbf{0}$ & $34.29$ & $\mathbf{0}$ \\
    \midrule
    Vehicle Dist. (km) $\downarrow$ & 47,096 & $\mathbf{26,946}$ &
    24,790 & $\mathbf{26,946}$ & 27,922 & $\mathbf{13,501}$ &
    $\mathbf{11,495}$ & 13,501 \\
    Package Dist. (km) $\downarrow$ & $52,950$ & $\mathbf{47,700}$ &
    $60,267$ & $\mathbf{47,700}$ & 31,385 & $\mathbf{22,979}$ &
    26,682 & $\mathbf{22,979}$ \\
    Dist. Saved (km) $\uparrow$ & 5,854 & $\mathbf{20,754}$ &
    $\mathbf{35,478}$ & 20,754 & 3,463 & $\mathbf{9,479}$ &
    $\mathbf{15,187}$ & 9,479 \\
    Dist. Saving (\%) $\uparrow$ & 11.1 & $\mathbf{43.5}$ &
    $\mathbf{58.9}$ & 43.5 & 11.0 & $\mathbf{41.2}$ & $\mathbf{56.9}$ & 41.2 \\
    \midrule
    Avg Time (sec) $\downarrow$ & 903 & $\mathbf{635}$ & 2,859 &
    $\mathbf{635}$ & 768 & $\mathbf{468}$ & 1,356 & $\mathbf{468}$ \\
    Median Time (sec) $\downarrow$ & 842 & $\mathbf{605}$ & 2,823 &
    $\mathbf{605}$ & 694 & $\mathbf{442}$ & 1,348 & $\mathbf{442}$ \\
    \midrule
    Bundles Created $\uparrow$ & 2,700 & $\mathbf{5,120}$ & 1,094 &
    $\mathbf{5,120}$ & 3,411 & $\mathbf{5,175}$ & 1,079 & $\mathbf{5,175}$ \\
    Bundling Part. $\uparrow$ & $\mathbf{9,234}$ & 7,604 &
    $\mathbf{9,234}$ & 7,604 & $\mathbf{9,234}$ & 7,684 &
    $\mathbf{9,234}$ & 7,684 \\
    \midrule
    Avg Delay (sec) $\downarrow$ & $\mathbf{-839}$ & -311 & $4$ &
    $\mathbf{-311}$ & $\mathbf{-1,000}$ & -254 & 4 & $\mathbf{-254}$ \\
    Median Delay (sec) $\downarrow$ & $\mathbf{-958}$ & -291 & 4 &
    $\mathbf{-291}$ & $\mathbf{-1,106}$ & -283 & 4 & $\mathbf{-283}$ \\
    Max Delay (sec) $\downarrow$ & 4,437 & $\mathbf{2,114}$ &
    $\mathbf{9}$ & 2,114 & 2,481 & $\mathbf{2,166}$ & $\mathbf{9}$ & 2,166 \\
    \bottomrule
  \end{tabular}
\end{table*}

\subsubsection{Comparison with MADDPG and GNN Approaches}
Table~\ref{tab:comparison_cities} compares UMST against MADDPG and
GNN baselines across Columbus and Chicago with 9,234 deliveries each.
Both baselines are implemented on UMST backbones. Results (tables and
graphs) for the baselines using the Clique backbone are provided in
the supplementary file and exhibit similar trends.

\noindent\textbf{Training Overhead.} UMST requires zero training
time, while MADDPG demands approximately 280-290 minutes, and GNN
requires 30-35 minutes per city for offline training. This 30× to
280× speedup in deployment readiness makes UMST immediately adaptable
to network changes without retraining costs. Training times for GNN
and MADDPG increase superlinearly with graph size: GNNs scale with
the number of edges ($\mathcal{O}(|E|)$ per layer), becoming
quadratic in dense graphs, while MADDPG exhibits exponential growth
in critic complexity with the number of agents.

\noindent\textbf{Success Rate and Reliability.} Performance varies
notably across methods. GNN achieves perfect 1.0 success in both
cities, while UMST attains 0.960 (Columbus) and 0.944 (Chicago),
outperforming MADDPG's 0.918 and 0.947 respectively. Completion rates
remain high (0.983 to 1.0) across all methods, indicating that
differences stem from time-constraint satisfaction rather than task abandonment.

\noindent\textbf{Operational Efficiency.} Distance efficiency shows
distinct patterns. GNN achieves highest distance savings (58.9\%
Columbus, 56.9\% Chicago) through aggressive consolidation, creating
only 1,094 and 1,079 bundles despite 100\% participation. UMST
maintains competitive savings (43.5\% and 41.2\%) while creating
substantially more bundles (5,120 and 5,175), enabling finer-grained
routing flexibility. MADDPG shows lower savings (11.1\% and 11.0\%)
with moderate bundle creation (2,700 and 3,411).

\noindent\textbf{Temporal Performance.} UMST achieves superior time
efficiency. Average delivery times for UMST (635.33 s Columbus,
468.42 s Chicago) significantly outperform GNN (2,858.95 s and
1,356.24 s) and remain competitive with MADDPG (903.40 s and 768.35
s). UMST maintains negative average delays ($-310.98$ s and $-254.39$
s), indicating consistent early arrivals, while GNN operates at
near-zero delays (4.43 s and 4.45 s), suggesting tighter time windows
with less operational buffer.

\noindent\textbf{Practical Implications.} UMST achieves competitive
performance across multiple objectives without requiring training.
While GNN attains perfect success rates, its dramatically longer
delivery times (4 to 6 times slower than UMST) limit practical
applicability for time-sensitive operations. MADDPG requires
extensive training and retraining when conditions change, whereas
UMST adapts immediately through simple graph reconstruction. These
characteristics make UMST particularly suitable for real-world
deployment where rapid adaptation, interpretability, and predictable
performance are essential.

\subsection{Extended Baseline Comparison: Attention-Based MADDPG}
\label{subsec:attention_maddpg}

This analysis extends the experimental evaluation by introducing an
Attention-Based MADDPG baseline. This model augments the standard
MADDPG critic with an attention mechanism \cite{iqbal2019actor},
allowing agents to dynamically weigh the importance of other agents'
observations during training. We compare this advanced learning-based
method against the vanilla MADDPG baseline and our proposed UMST approach.

\begin{table*}[t]
  \centering
  \caption{Comprehensive performance comparison: Vanilla MADDPG vs.
    Attention-Augmented MADDPG vs. UMST ($N=9,234$). Attention times
  converted to seconds for consistency.}
  \label{tab:performance_comparison}
  \vspace{0.2cm}
  \resizebox{\textwidth}{!}{
    \begin{tabular}{l|ccc|ccc}
      \toprule
      & \multicolumn{3}{c}{\textbf{Columbus}} &
      \multicolumn{3}{c}{\textbf{Chicago}} \\
      \textbf{Metric} & \textbf{MADDPG} & \textbf{Attn-MADDPG} &
      \textbf{UMST} & \textbf{MADDPG} & \textbf{Attn-MADDPG} & \textbf{UMST} \\
      \midrule
      Total Deliveries & 9,234 & 9,234 & 9,234 & 9,234 & 9,234 & 9,234 \\
      Completed Del. $\uparrow$ & 9,234 & 9,234 & 9,103 & 9,234 &
      9,234 & 9,080 \\
      Successful Del. $\uparrow$ & 8,472 & 8,979 & 8,868 & 8,740 &
      9,102 & 8,720 \\
      Failed Del. $\downarrow$ & 762 & 254 & 235 & 494 & 132 & 360 \\
      Success Rate $\uparrow$ & 0.918 & 0.972 & 0.960 & 0.947 & 0.986 & 0.944 \\
      Completion Rate $\uparrow$ & 1.0 & 1.0 & 0.986 & 1.0 & 1.0 & 0.983 \\
      Training Time (min) $\downarrow$ & 293.09 & 983.34 & 0.00 &
      276.63 & 1,107.70 & 0.00 \\
      \midrule
      Vehicle Dist. (km) $\downarrow$ & 47,096 & 39,199 & 26,946 &
      27,922 & 17,306 & 13,501 \\
      Package Dist. (km) $\downarrow$ & 52,950 & 41,144 & 47,700 &
      31,385 & 18,795 & 22,979 \\
      Dist. Saved (km) $\uparrow$ & 5,854 & 1,947 & 20,754 & 3,463 &
      1,487 & 9,479 \\
      \midrule
      Avg Time (sec) $\downarrow$ & 903 & 550 & 635 & 768 & 307 & 468 \\
      Median Time (sec) $\downarrow$ & 842 & 356 & 605 & 694 & 255 & 442 \\
      \midrule
      Avg Delay (sec) $\downarrow$ & -839 & -1,161 & -311 & -1,000 &
      -1,441 & -254 \\
      Median Delay (sec) $\downarrow$ & -958 & -1,444 & -291 & -1,106
      & -1,545 & -283 \\
      Max Delay (sec) $\downarrow$ & 4,437 & 5,281 & 2,114 & 2,481 &
      2,159 & 2,166 \\
      \bottomrule
    \end{tabular}
  }
\end{table*}

\subsubsection{Comparative Results}
Table \ref{tab:performance_comparison} presents a comprehensive
comparison across Columbus and Chicago. All time-based metrics for
the Attention baseline have been converted from minutes to seconds to
ensure direct comparability.

Key observations include:
\begin{itemize}
  \item \textbf{Reliability:} The Attention mechanism improves
    success rates over Vanilla MADDPG (rising from 0.918 to 0.972 in
    Columbus), effectively matching the reliability of UMST.
  \item \textbf{Efficiency:} UMST still achieves superior distance
    savings compared to other baselines (e.g., 26,946 km vs 39,199 km
    for Attention in Columbus).
  \item \textbf{Computational Cost:} The primary drawback of the
    attention mechanism is scalability. Training time jumps from
    approx. 293 minutes (Vanilla) to over 983 minutes (Attention),
    whereas UMST requires zero training time.
\end{itemize}

The inclusion of attention mechanisms significantly mitigates the
coordination failures observed in vanilla MADDPG. However, this comes
at the cost of a 3x increase in training time and does not reach the
routing efficiency of the structural UMST approach. UMST remains the
most balanced solution, offering competitive reliability with orders
of magnitude faster deployment.

\section{Conclusion, Limitations, and Future Work}
\label{sec:conclusion}

This paper presents the Union Minimum Spanning Tree (UMST) as a simple yet effective backbone for urban delivery networks. UMST is designed around a clear principle: good delivery systems need to be efficient without being fragile. By combining multiple lightly perturbed minimum spanning trees, UMST retains the low-cost nature of sparse graphs while introducing enough structural diversity to support reliable routing and order bundling. This balance allows the system to operate efficiently under normal conditions and remain functional during disruptions.

Our empirical evaluation shows that UMST consistently occupies the middle ground between two extremes: the rigidity of a single MST and the inefficiency of fully connected graphs. UMST supports high success rates, moderate delivery times, and substantial reductions in vehicle distance, particularly when bundling is enabled. Importantly, UMST achieves these gains through graph design alone, without relying on complex optimization or learning procedures.

When compared with learning-based approaches such as Graph Neural Networks (GNNs) and Multi-Agent Deep Deterministic Policy Gradient (MADDPG), UMST highlights a different but complementary philosophy. While learning-based methods can achieve high success rates, they require extensive training, careful tuning, and retraining when conditions change. In contrast, UMST requires no training. It adapts immediately to new demand patterns or network disruptions through fast graph reconstruction. Despite operating on a reduced backbone rather than the full road network, UMST delivers competitive or better performance in delivery time and travel distance, making it particularly attractive for time-sensitive and resource-constrained deployments.

UMST also offers practical advantages that are often critical in real-world systems. Its deterministic construction produces interpretable routing structures that support debugging, analysis, and human oversight. The ability to rapidly reconstruct the backbone in response to road closures, congestion, or demand shifts makes UMST naturally resilient, without the latency and uncertainty associated with retraining learning-based models.

There are several directions for future work. Our current implementation applies uniform perturbation during tree generation and relies on greedy bundling decisions at hotspots. Future extensions could incorporate spatially adaptive perturbations and time-varying edge weights to reflect traffic dynamics.
Extending UMST to heterogeneous fleets and integrating lightweight online learning for highly volatile demand also remain promising directions.

In conclusion, UMST demonstrates that careful graph design can deliver many of the benefits attributed to complex learning systems while avoiding their overhead, opacity, and adaptation challenges. By explicitly balancing sparsity and redundancy, UMST provides a robust, scalable, and interpretable foundation for urban delivery operations in dynamic environments.

\bibliography{ref}




\end{document}